\documentclass[pre,twocolumn,showpacs,preprintnumbers,amsmath,amssymb,superscriptaddress]{revtex4}

\usepackage{graphicx}
\usepackage{dcolumn}

\newcommand{\be}{\begin{equation}}
\newcommand{\ee}{\end{equation}}
\newcommand{\bd}{\begin{displaymath}}
\newcommand{\ed}{\end{displaymath}}
\newcommand{\BE}{\begin{eqnarray}}
\newcommand{\EE}{\end{eqnarray}}

\newcommand{\erf}{{\rm erf}}

\newcommand{\bx}{\ensuremath{\mathbf{x}}}

\newcommand{\avg}[1]{\left\langle{#1}\right\rangle}

\begin{document}


\title{Rank abundance relations in evolutionary dynamics of random replicators}
\author{Yoshimi Yoshino}
\email{yoshimi@cp.cmc.osaka-u.ac.jp}
\affiliation{Graduate School of Science and Cybermedia Center, Osaka University, Toyonaka, Osaka 560-0043, Japan}
\author{Tobias Galla}
\email{tobias.galla@manchester.ac.uk}
\affiliation{The Abdus Salam International Centre for Theoretical Physics, Strada Costiera 11, 34014 Trieste, Italy}
\affiliation {The University of Manchester, School of Physics and Astronomy, Schuster Building, Manchester M13 9PL, United Kingdom}
\author{Kei Tokita}
\email{tokita@cmc.osaka-u.ac.jp}
\affiliation{Graduate School of Science and Cybermedia Center, Osaka University, Toyonaka, Osaka 560-0043, Japan}
\affiliation{Graduate School of Frontier Biosciences, Osaka University, Suita, Osaka 565-0871, Japan}

\date{\today}

\begin{abstract}
We present a non-equilibrium statistical mechanics description of rank
abundance relations (RAR) in random community models of ecology.
Specifically, we study a multi-species replicator system with quenched
random interaction matrices. We here consider symmetric interactions
as well as asymmetric and anti-symmetric cases. RARs are obtained
analytically via a generating functional analysis, describing
fixed-point states of the system in terms of a small set of order
parameters, and in dependence on the symmetry or otherwise of
interactions and on the productivity of the community. Our work is an
extension of Tokita [Phys. Rev. Lett.  {\bf 93} 178102 (2004)], where
the case of symmetric interactions was considered within an
equilibrium setup. The species abundance distribution in our model
come out as truncated normal distributions or transformations thereof
and, in some case, are similar to left-skewed distributions observed
in ecology. We also discuss the interaction structure of the resulting
food-web of stable species at stationarity, cases of heterogeneous
co-operation pressures as well as effects of finite system size and of
higher-order interactions.
\end{abstract}
\pacs{PACS}

\maketitle


\section{Introduction}

Understanding the relationship between complexity and stability is a
fundamental and controversial problem in ecology \cite{McCann_2000}.
Before the 1970s the proposition that highly complex communities are
more stable than simple ones was widely supported
\cite{MacArthur_1955, Elton_1958}. However, this early intuitive idea
was challenged by theorists in the 1970s, who discussed the stability
of a community of species interacting randomly
\cite{Gardner_Ashby_1970}. In particular, the applications of random
matrix theory rigorously revealed that the stability of a community
strongly depends on complexity, e.g. diversity and statistical
properties of the interaction matrix, such as variance and
connectivity, and complexity tends to destabilize community dynamics
\cite{May_1972}.  Since then, many mathematical ecologists have
studied random community models to explain the apparent contradiction
between the complexity of real-world ecosystems and the results of
these mathematical studies
\cite{May_1974,Pimm_1991}. Recent theoretical developments, for example, have
discovered stabilizing factors of random community models: competition
\cite{Rozdilsky_Stone_2001} and antisymmetric prey-predator
relationships \cite{Chawanya_Tokita_2002}. Empirical and theoretical
works also suggested importance of omnivory (higher connectance)
\cite{Lawler_Morin_1993, McCann_Hastings_Huxel_1998} and weak
interactions \cite{Neutel_etal_2002} for stability.

If the relative abundances of the species in a community are measured,
inevitably a small number of very common species will be identified
(i.e. species with a high abundance), along with some rare species and
more numerous species of varying intermediate degrees of rareness.
Clarifying the mechanisms underlying these rank abundance relations (RAR)
(the relations between abundance and the number of species possessing
that abundance) are clearly another fundamental problem of ecology
\cite{Brown_1995,Rosenzweig_1995}. In conservation biology as well,
knowledge of RAR helps one to predict the
likelihood of population persistence and community stability in face
of global change. Various models have been applied to ecosystem
communities
\cite{May_1975,Sugihara_1980,Nee_Harvey_May_1991,Tokeshi_1998,Hall_etal_2002}
and, in special, the recent progress of the theory of `neutral' models 
\cite{Hubbell_2001,Volkov_etal_2003,Alonso_McKane_2004,Etienne_Olff_2004,Etienne_2005,Alonso_etal_2006, Etienne_Alonso_2007} have aroused constructive discussions on theoretical predictions and the experimental studies on RAR. As the neutral models mainly cover ecosystem communities
where species compete for niches on a single trophic level like a tropical forest or a coral reaf, the models have left the more complex systems a mystery. 
Such systems occur on multiple trophic levels and include complex
interactions, such as prey-predator relationships, mutualism,
competition, and detritus food chains. Although RAR are observed 
universally in nature, their essential
parameters have not been fully clarified. 

As a step to explore RAR
theoretically, in this paper, typical rank abundance relations are
derived using a random community model with few parameters such as the
level of symmetry of interaction matrix and co-operation pressure or
productivity. While random community models can be criticized for a
lack of immediate realism, they have the advantage of being exactly
solvable by analytical techniques. Random replicator systems have for
example been considered as solvable models of interacting species in
\cite{Diederich_Opper_1989, Opper_Diederich_1992,Tokita_2004}. In particlar, species
abundance distributions of random replicator models with symmetric
couplings have been computed in \cite{Tokita_2004} using methods from
equilibrium statistical mechanics. Such static approaches are limited
to cases of symmetric couplings between species, in particular the
presence of predator-prey pairs (for which interactions are highly
asymmetric) can not be taken into account in such equilibrium
approaches. In order to remedy these shortcomings, we here take a
different dynamical approach, allowing for an extension to systems
with an arbitrary proportion of predator-prey pairs. To this end we
employ methods different from those of \cite{Tokita_2004} and focus on
an approach based on dynamical generating functionals and path
integrals.

It is interesting to note that stochastic models of complex
dynamically assembled food-webs
\cite{McKane_Alonso_Sole_2000,Sole_Alonso_McKane_2002}, which is from a simple dynamics
governed by generalized birth and death events, derive reasonable species
abundance distributions, in good agreement with real data. In such
models the multi-species dynamics is effectively reduced to that of a
representative species, subject to a 'mean field' interaction with the
remaining system. In a similar fashion our approach reduces the evolution
of species randomly coupled via quenched interactions to a
`one-species' effective process as well (albeit a non-Markovian
one). This mapping leads to an exact solution in the thermodynamic
limit of infinite system size. 
For the stochastic approach, the model has the randomness with some
probability. On the other hand, for generation functional, it gives the
fixed randomness in the deterministic time evolution. 
Apart from providing a starting point
for more realistic modifications of the present model, our analysis
can hence, to a certain degree, be seen as complementary to the
approach of \cite{McKane_Alonso_Sole_2000,Sole_Alonso_McKane_2002}.

In the context of statistical mechanics another interesting point of
the present model is that the replicator dynamics with asymmetric
random interactions shows a non-equilibrium phase transition, i.e. two
phases with qualitatively different behaviors are found (stable
versus unstable). At the same time the replicator system does not
exhibit a Lyapunov function, and is hence intrinsically a
non-equilibrium model without detailed balance. Further details can be
found in the statistical mechanics literature \cite{Mazenko_2006,Pathria_1996}. In our system
destabilization of a globally fixed point solution and its bifurcation
to limit cycle, heteroclinic cycle and potentially chaos is found when
parameters are varied. The random replicator model hence shows
similarities, but also crucial differences compared with e.g. models
of spin glasses \cite{Sherrington_Kirkpatrick_1978,Mezard_etal_1987}
and neural network models \cite{During_etal_1998,Mimura_2004}. It is
hoped that the study of random replicator dynamics may hence contribute to the understanding and classification of dynamical phase transitions in disordered systems.

This paper is organized as follows: we will define the model in Sec.
II and then discuss the statistical mechanics analysis based on a
path-integral approach in Sec. III. In Sec. IV, we show results for
pairwise interaction: a stability analysis, phase diagram, survival
function, rank-abundance relations (RAR), the species abundance
distribution (SAD), finite size effects and structure of the resulting
food web are discussed. We then turn to heterogeneous co-operation
pressure and higher-order interactions in subsequent sections V and
VI, respectively. We summarize our results in Sec.VII.

\section{Model}
We here study the simplest system of random
replicator subject to Gaussian interaction, and focus on the model
originally proposed by Diederich and Opper
\cite{Diederich_Opper_1989, Opper_Diederich_1992}. In conventional replicator dynamics, the system consists
of $N$ species, labeled by $i=1,\dots,N$. The composition of the
population of species at time $t$ is then described by a concentration
vector $\bx(t)=(x_1(t),\dots,x_N(t))$, where $x_i(t)$ denotes the
concentration of species $i=1,\dots,N$, and where $\sum_i
x_i(t)=1$. The system evolves in time
according to the following replicator equations
\cite{Hofbauer_Sigmund_1988}
\be\label{eq:repl}
\frac{\dot x_i(t)}{x_i(t)}=f_i[\bx(t)]-\nu(t),
\ee
where $f_i[\bx]$ is the `fitness' of species $i$ at time $t$, and where $\nu(t)$ denotes the mean fitness of species in the population. Hence species fitter than average increase in concentration, whereas the weight of species less fit than average is reduced.

We here take the fitnesses $f_i[\bx]$ to be frequency-dependent, i.e. they are functions of the vector $\bx$. Specifically we will assume, in the simplest setting, that
\be\label{eq:fit}
f_i[\bx]=-2u x_i+\sum_{j\neq i} w_{ij} x_j,
\ee
i.e. that interaction between species is pairwise and characterized by
the matrix elements $w_{ij}$. Generalization to multi-species
interaction is possible \cite{de_Oliveira_Fontanari_2000, Galla_2006a}, and will be discussed below.

The matrix elements $\{w_{ij},w_{ji}\}$ (for any pair $i<j$) are
chosen from a Gaussian ensemble. Specifically we choose \be
\overline{w_{ij}}=0, ~~~ \overline{w_{ij}^2}=\frac{w^2}{N}, ~~~
\overline{w_{ij}w_{ji}}=\Gamma\frac{w^2}{N}, \ee where
$\overline{\cdots}$ denotes an average over the random couplings.  $w$
here characterizes the magnitude of the interaction, and $\Gamma$ is a
symmetry parameter and takes values $\Gamma\in [-1,1]$. For $\Gamma=1$
the interaction between any pair of species $i<j$ is fully symmetric,
$w_{ij}=w_{ji}$. In this case no predator-prey pairs are found in the
system. For $\Gamma=0$ $w_{ij}$ and $w_{ji}$ are uncorrelated, the fraction of predator-prey pairs is hence $50$ per cent. For $\Gamma=-1$ all pairs of species are in predator-prey constellations, one here has
$w_{ij}=-w_{ji}$. Choosing intermediate values of $\Gamma$ allows one
to interpolate smoothly between these regimes. The ecologically most
relevant setup corresponds to negative values of $\Gamma$, describing
prey-predator type interaction between species, rather than
co-operation and direct mutual competition.  Diagonal terms in
Eq. (\ref{eq:fit}) can be taken into account by writing $w_{ii}=-2u$,
where $u$ in the above setting denotes the so-called co-operation
pressure \cite{Paschel_Mende_1986}. In an ecological context $u$ takes
mostly positive values. For $u\to\infty$ the ecosystem is found in a
state of perfect co-operation and maximal diversity (with all species
surviving and having equal concentrations). The essential parameter
$p=2u$ can be termed as the productivity of a community in the sense
of Lotka-Volterra equation (this will be explained in more detail in
Sec. IV). Finally, in order to guarantee a well-defined thermodynamic
limit $N\to\infty$, with which the statistical mechanics analysis of
the model will be concerned, we re-scale the concentration vector by a
factor of $N$, and use the normalization $N^{-1}\sum_i x_i(t)=1$. Upon
setting $\nu(t)=N^{-1}\sum_i x_i(t) f_i[\bx(t)]$ this normalization is
conserved by the replicator dynamics (\ref{eq:repl}).

We will address the model by a combination of analytical and computational methods. The statistical mechanics theory is described in the following sections, and its results will be compared against simulations in the subsequent section. All simulations are here performed using the method described in \cite{Opper_Diederich_1999}. This numerical scheme effectively amounts to a first-order forward integration with a dynamically adapted time-step. The latter is here necessary to avoid species concentrations to go negative in the discretized system. The dynamical time-stepping used in our simulations if typically of the order of $0.01$ to $0.1$.

\section{Statistical mechanics theory}\label{sec:gfa}
\subsection{Path integral analysis}
The above system can be addressed by generating functional techniques
originally devised in the theory of disordered systems
\cite{DeDominicis_1978}. It is also applied to linear 
evolutionary dynamics model in \cite{Rieger_1989} and can be adapted to the study of random Lotka-Volterra communities \cite{Galla_forthcoming}.
We will not detail the mathematical steps
here, as they have been reported in depth in the literature
\cite{Opper_Diederich_1992,Galla_2006a}. In the thermodynamic limit the system is
found to be described by an effective single-species process of the
form \cite{Opper_Diederich_1992}
\be
\dot x(t)=x(t)\bigg(-2ux (t)-\Gamma \int_{t_0}^t dt' G(t,t') x(t')-\eta(t)-\nu(t)\bigg)\label{eq:effproc}
\ee
($t_0$ denotes the time at which the dynamics is started). This process is non-Markovian in
time, and subject to colored Gaussian noise $\eta(t)$, with temporal
correlations given by
\be
\avg{\eta(t)\eta(t')}=C(t,t').\label{eq:effproc_eta}
\ee
This colored noise is obtained from the interactions of randomness,
which each trajectory has.
$C(t,t')$ and $G(t,t')$ are the correlation
and response functions, and are to be evaluated self-consistently as
\be
C(t,t')=\avg{x(t)x(t')}_\star, ~~ G(t,t')=\avg{\frac{\delta x(t)}{\delta \nu(t')}}_\star,\label{eq:effproc_CG}
\ee
where $\avg{\cdot}_\star$ denotes an average over trajectories of the
effective stochastic process (\ref{eq:effproc}).  The analysis then
proceeds by making a fixed-point ansatz, amounting to $Q \equiv
C(t,t')$. We also write $\chi=\int dt G(t)$ for the integrated
response, and consider only ergodic states in which $\chi$ remains
finite. Restricting the analysis to asymptotically time-independent
solutions of the effective process the following self-consistent
equations for the resulting static order parameters $Q,\chi$ and $\nu$
(the fixed-point value of the average fitness) can then be derived
similar to those reported in \cite{Opper_Diederich_1992}
\BE
\frac{M}{\sqrt{\lambda}}&=&\int_{-\infty}^{\Delta} Dz (\Delta-z),\label{eq:firsteq}\\
\frac{Q M^2}{\lambda}&=&\int_{-\infty}^{\Delta} Dz (\Delta-z)^2,\label{eq:secondeq}\\
-M \chi&=&\int_{-\infty}^\Delta Dz.\label{eq:thirdeq}
\EE
Here $Dz=\frac{1}{\sqrt{2\pi}}e^{-z^2/2}dz$ denotes the standard
Gaussian measure, and one has 
$\lambda=w^2Q$, 
$M =2u+w^2 \Gamma \chi $
and
$\Delta=-\nu/\sqrt{\lambda}$. We note that $\phi=\int_{-\infty}^\Delta
Dz=\frac{1}{2}\left(1+\erf\left(\Delta/\sqrt{2}\right)\right)$ describes
the fraction of surviving species. These equations are readily solved numerically, providing analytical predictions of the statistics of fixed-point solutions as functions of the model parameters $w,\Gamma$ and $u$. 

\subsection{Co-operation pressure and strength of interaction}
\begin{figure}[t]
\centerline{\includegraphics[width=0.35\textwidth]{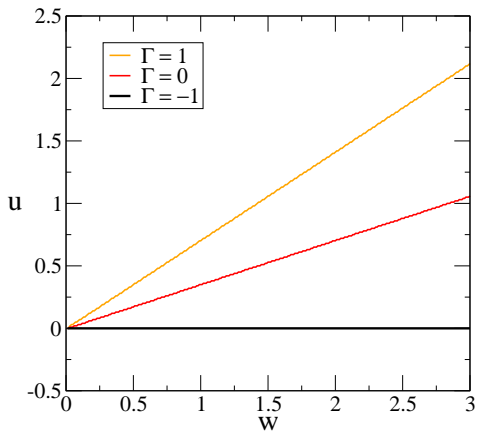}}
\caption{(Color online) Phase diagram of the model with pairwise interaction in the $(w,u)$ plane for $\Gamma=1, 0, -1$ from top to bottom. The system approaches a unique stable fixed point in the region above the respective lines, and remains unstable and non-ergodic below the phase boundaries.}
\label{fig:pd_asym}
\end{figure}

For reasons of completeness we re-iterate the phase behavior of the model as obtained by a linear stability analysis first reported in \cite{Opper_Diederich_1992}. One here finds a stable region in which the fixed-point of the replicator dynamics is unique and locally attractive, separated from an unstable phase, as shown in Fig. \ref{fig:pd_asym}.

For $\Gamma=-1$, the system is always found to be stable for any $u>0$
independently of $w$.  At fixed $w=1$, the onset of instability occurs
at $u_c = \sqrt{2}/4$ and $u_c = \sqrt{2}/2$ for $\Gamma=0$ and
$\Gamma=1$ respectively. While the generating functional approach is
applicable for general symmetry parameter $\Gamma$, a static analysis
based on the replica method is possible for symmetric couplings
($\Gamma=1$). This has been carried out in
\cite{Diederich_Opper_1989, Biscari_Parisi_1995}. The replica approach is here fundamentally different from ours,
as it is only of a static (time-independent) nature. The ergodic
fixed-point phase corresponds to a regime in the static analysis in
which only one well-defined minimum of the Lyapunov function is found,
corresponding to a so-called replica symmetric solution \cite{Mezard_etal_1987}. This solution becomes unstable at
the phase transition, referred to as a de Almeida-Thouless
instability, coinciding with the location dynamical instability has
been identified. For $u_c < u_c(\Gamma=1,w=1)=\sqrt{2}/2$ replica
symmetry breaking (RSB) occurs, i.e. the manifold of minima of the
Lyapunov function becomes disconnected \cite{Mezard_etal_1987}. In conclusion, while the replica approach, requiring the existence of a Lyapunov function, is limited to the case $\Gamma=1$, generating functionals can be used to study the replicator system for any degree of asymmetry in the interaction matrix, as this approach requires only the knowledge of the dynamical equations of the system (the replicator equations), but is independent of the existence or otherwise of a quantity minimized by these dynamics. For the case of symmetric couplings, $\Gamma=1$, the results from both methods coincide.

Finally, since the behavior of the system can be seen to be
qualitatively independent of the coupling strength $w$ (which
effectively re-scales the co-operation pressure), we will focus solely
on $w=1$ in the following.

\subsection{Species Abundance Distribution (SAD)}

Making a fixed point ansatz in the effective process
(\ref{eq:effproc}) amounts to considering the time-independent
solution of the effective species process of the form
\cite{Opper_Diederich_1992} \be\label{eq:sol}
x(z)=\frac{-\nu-\sqrt{\lambda}z}{M}\Theta\left(-\nu-\sqrt{\lambda}z\right),
\ee which represents the stochastic expression of the population in
the stable state.  $z$ is here a static random variable drawn from a
standard Gaussian distribution ($\sqrt{\lambda}z$ reflects the
single-particle noise $\eta(t)$ which becomes time-independent in the
fixed-point regime). $\Theta(\cdot)$ is the step function. Note that,
as mentioned above, only a fraction of species have positive
concentrations at the fixed point, and that a complementary fraction
of species dies out asymptotically. The distribution of concentrations
$x$ at the fixed point is thus a Gaussian cut-off at $x=0$ combined
with a delta-peak at $x=0$ \cite{Opper_Diederich_1999}. The so-called
{\em survival function} \be \alpha(x)=\lim_{N \to
\infty}\frac{1}{N}\sum_i \overline{\Theta\left(x_i-x\right)}, \ee
denotes the fraction of species with a concentration strictly larger
than $x$ at the fixed point. The survival function, indicating the
probability of a species having an abundance larger than $x$, is
easily computed from (\ref{eq:sol}) and is found as \be
\alpha(x)=\frac{1}{2}\left(1+\erf\left(\frac{\Delta-\frac{M}{\sqrt{\lambda}}x}{\sqrt{2}}\right)\right)
\ee in the thermodynamic limit. The fraction of survivors $\phi$ as
defined above is obtained as the special case $\phi=\alpha(x=0)$.

Using the cumulative distribution function $C(x) \equiv 1-\alpha(x)$
(denoting the probability for a species to have a concentration less
than or equal to $x$) the abundance distribution for $x>0$ is given by
\be
F(x)=\frac{dC(x)}{dx}=\frac{M}{\sqrt{2\pi\lambda}} \exp\left(-\frac{(
\Delta-\frac{M}{\sqrt{\lambda}}x)^2}{2}
\right)
\ee
A similar expression has been obtained for the case of symmetric couplings ($\Gamma=1$) based on replica techniques in \cite{Tokita_2004}. These earlier findings are found from our generating functional analysis as a limiting case,
so that generation functional analysis contains the technique of replica method as mentioned above.

\section{Results for pairwise interaction}

\subsection{Survival function}
\begin{figure}[t]
\vspace{3em}
\centerline{\includegraphics[width=0.35\textwidth]{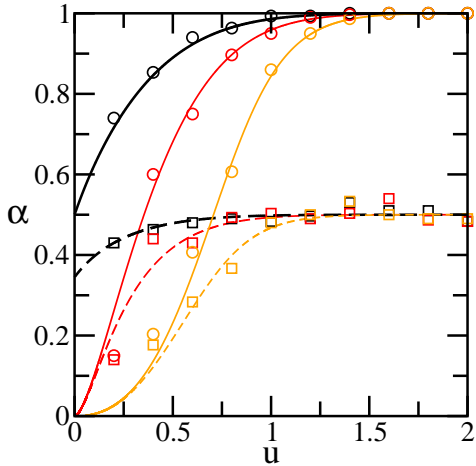}}
\caption{(Color online) Survival functions $\alpha(0)$ and $\alpha(1)$ as functions of the co-operation pressure at fixed $w=1$. Upper curves show $\alpha(0)$, lower curves $\alpha(1)$, with $\Gamma=-1,0,1$ from top to bottom in each group. Lines are from theory (valid only above $u_c(\Gamma)$), symbols from simulations of systems with $N=300$ species, averaged over $20$ samples. Surviving species in simulations are identified as species with $x_i>0.01$ asymptotically.}
\label{fig:survival}
\end{figure}

We plot the survival functions $\alpha(x=0)$ and $\alpha(x=1)$ as a
function of the co-operation pressure and for different values of
$\Gamma$ in Fig. \ref{fig:survival}. As seen in the figure the
diversity of the population (as measured for example by the number of
surviving species) increases with larger co-operation pressure.  The
figure also demonstrates good agreement between numerical simulations
and theoretical predictions for large values of the co-operation
pressure $u$. In this phase the system is stable and ergodic and hence
the fixed-point theory applies. Numerical simulations are performed
using the discretization scheme described in
\cite{Opper_Diederich_1999}. Below a critical value $u_c(\Gamma)$
stability and ergodicity are lost (for $\Gamma>-1$), and the above
theory can no longer be expected to be accurate, and systematic
deviations between theory and simulations may occur. Still the
qualitative agreement between theoretical lines, extended into the
unstable phase, where they are technically no longer valid, is
surprisingly good (RSB effects have been seen to be weak in the low-$u$ phase in previous studies). No unstable
phase is present for fully anti-correlated couplings ($\Gamma=-1$) and
non-negative co-operation pressure.

\subsection{Rank-abundance relations}
If the $S=\phi N$ surviving species are re-labeled and ordered according to their abundance in descending order, i.e. if $x_1\geq x_2\geq ...\geq x_{S}$ then $\alpha(x)$ can be understood as representing the species rank $n$ according to
\BE
\alpha (x) = \frac{n}{N} \; \mbox{for} \; x \in [x_{n+1}, x_n).
\EE
The function $\alpha(x)$ is a non-increasing monotonic function, and can hence be inverted. The abundance $x(n/N)$ of the $n$-th most abundant species can then be written as
\BE
x(n/N)=\alpha^{-1}(n/N).
\EE
This representation is generally referred to as a `rank abundance relations' (RAR)
in the ecology literature. We find typical sigmoidal patterns which
have been observed in different regions \cite{Hubbell_2001} and with
different species compositions \cite{Whittaker_1970}.  In general, for
large value of $u$ the RAR are broad and corresponds to RAR for a
species-rich community. Remarkably, the cross-over of the RAR patterns
from low- to high- $u$ is similar to the observed transition from
low- to high productivity areas in real-world data, that is, comparing species-poor
areas such as an alpine or polar region to a species-rich tropical rain
forest \cite{Hubbell_2001}. The transition also corresponds to the
secular variation of patterns observed in abandoned cultivated land
\cite{Bazzaz_1975}. This supports the contention that $u$ is a maturity
parameter, as is suggested by an earlier evolutionary model in 
\cite{Tokita_Yasutomi_2003}. 

\begin{figure}[t]\vspace{3em}
\centerline{\includegraphics[width=0.35\textwidth]{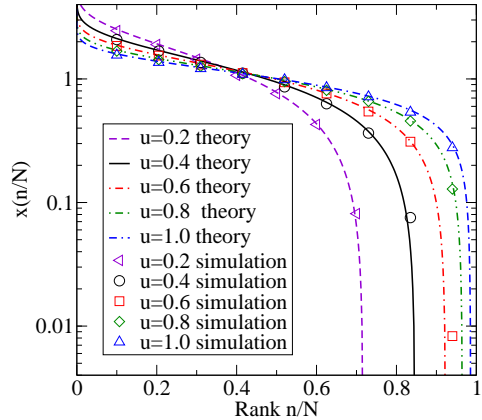}}
\caption{(Color online) Rank abundance relation for $w=1$, $\Gamma=-1$. Markers are from simulations. ($N=200$, $20$ samples, $10000$
 iterations using the integration scheme of \cite {Opper_Diederich_1999}), lines from the fixed point theory.}
\label{fig:a-1_G-1}
\end{figure}

\begin{figure}[t]\vspace{3em}
\centerline{\includegraphics[width=0.35\textwidth]{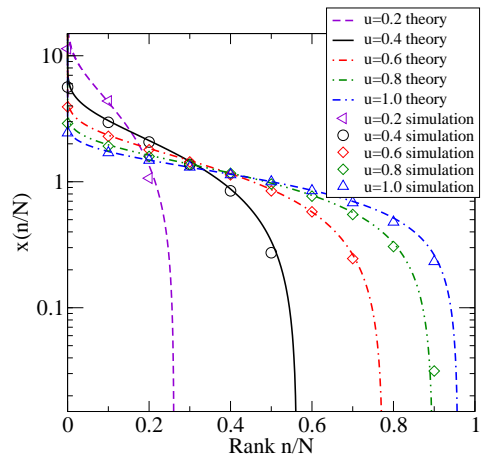}}
\caption{(Color online) Rank abundance relation for $w=1$. $\Gamma=0$. Markers are from simulations. ($N=200$, $20$ samples), lines from the fixed point theory, valid for $u>u_c=\sqrt{2}/4$, and of an approximate nature for $u<u_c$.}
\label{fig:a-1_G0}
\end{figure}

\begin{figure}[t]\vspace{3em}
\centerline{\includegraphics[width=0.35\textwidth]{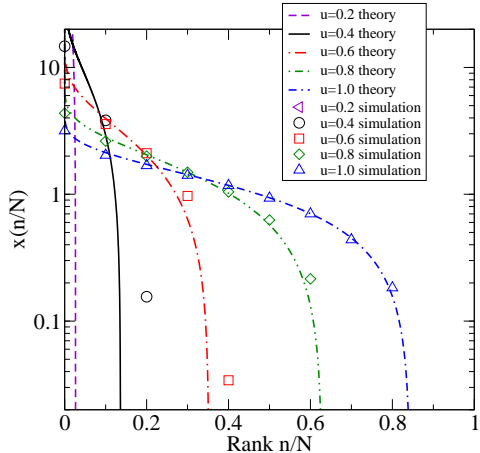}}
\caption{(Color online) Rank abundance relation for $w=1$, $\Gamma=1$. Markers are from simulations. ($N=200$, $20$ samples), lines from the fixed point theory, valid for $u>u_c=\sqrt{2}/2$, and of an approximate nature for $u<u_c$.}
\label{fig:a-1_G1}
\end{figure}

\subsection{Species abundance distribution and Preston's octave plot}
\begin{figure}[t]\vspace{3em}
\centerline{\includegraphics[width=0.35\textwidth]{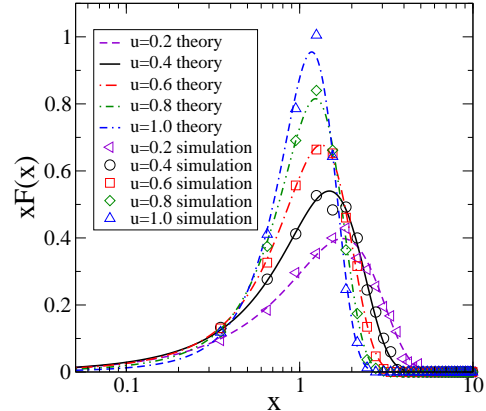}}
\caption{(Color online) $\Gamma = -1$. The lines are from theory, $u=1.0, 0.8, 0.6,
 0.4, 0.2$ from top to bottom. The stable phase extends to all $u>0$. Markers are from simulations.($N=200$, averages over $50$ samples are taken).}
\label{fig:xFx_G-1}
\end{figure}

\begin{figure}[t]\vspace{3em}
\centerline{\includegraphics[width=0.35\textwidth]{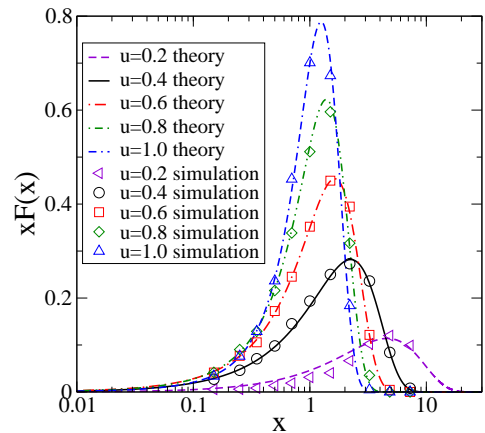}}
\caption{(Color online) $\Gamma = 0$. The lines are from theory, $u=1.0, 0.8, 0.6,
 0.4, 0.2$ from top to bottom. Stable phase contains $u=1.0, 0.8, 0.6,
 0.4$. $u=0.2$ is in the unstable phase where the theory applies only as
 an approximation ($u_c=\sqrt{2}/4$), Markers are from simulations. ($N=200$, $50$ samples). }
\label{fig:xFx_G0}
\end{figure}

\begin{figure}[t]\vspace{3em}
\centerline{\includegraphics[width=0.35\textwidth]{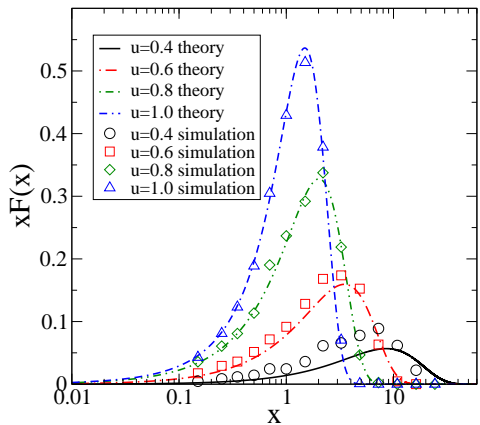}}
\caption{(Color online) $\Gamma = 1$. The lines are from theory, $u=1.0, 0.8, 0.6,
 0.4$ from top to bottom. Stable phase $u=1.0, 0.8$ and unstable phase $u=0.6, 0.4$ ($u_c = \sqrt{2}/2$). Markers are from simulations. ($N=200$, $50$ samples).}
\label{fig:xFx_G1}
\end{figure}

Empirical data of species abundance have been taken for example in the studies of \cite{Fisher_Corbet_Williams_1943,
Preston_1962a, Preston_1962b, MacArthur_1957}, and are normally presented as plots of `species per octave'. I.e. species are grouped according to their abundance, and any species with abundance (number of individuals of that species present in the eco-system) in the interval of say $[2^n,2^{n+1})$ is subsumed in octave $n$ ($n$ being an integer). Log-normal distribution are then observed e.g. in \cite{Preston_1962a,
Preston_1962b}. In order to depict the species abundance distributions in a manner similar to
Preston's octave plot, we plot $xF(x)$ versus $x$ in a log scale following \cite{Tokita_2004}, see Figs.  
\ref{fig:xFx_G-1}, \ref{fig:xFx_G0} and \ref{fig:xFx_G1} \footnote{We
here note that upon writing $\omega_i=\ln(x_i)$ for the octave to which
a species with concentration $x_i>0$ belongs, and $G(\omega)$ for the
`density of species' per octave $\omega$, one has
$\int_{\omega}^{\omega+1}G(\omega')d\omega'=\int_{e^\omega}^{e^{\omega+1}}F(x)dx$,
so that one realizes by substitution that $G(\omega)=e^\omega
F(e^\omega)=xF(x)$, which motivates our plotting of $xF(x)$ in a
log-linear scale. Note that for convenience we do not use a base of two
in this context, but choose natural logarithms instead. This amounts to no  
more than an overall re-scaling by a constant factor.}.

Generally, we find that an increased co-operation pressure
(equivalently an increased productivity, see below) larger $u$ leads
to `octave plots' with small average and small variance. Species
concentrations are here mostly found at a value of around $x=1$ (in
the limit of infinite co-operation pressure, $u\to\infty$, all species
have equal concentrations), and hence it is mostly the octave
containing $x=1$ which is populated. On the other hand, for smaller
$u$, fewer species survive, and the variance in their concentrations
can be significant. This leads to octave plots of a large variance and
a left-skewed form, similar to shapes observed e.g. in
\cite{Nee_Harvey_May_1991,Hubbell_2001}. In the fully asymmetric case $\Gamma=-1$, see
Fig. \ref{fig:xFx_G-1} all theoretical curves are in good agreement
with results from simulations for all values of $u$. Here the theory
is exact. In Figs. \ref{fig:xFx_G0} and \ref{fig:xFx_G1}, however,
corresponding to $\Gamma=0$ and $\Gamma=1$ the theory is valid only
for $u>u_c(\Gamma)$. Good agreement between analytics and simulations
is again observed. For $u<u_c$ the theory is at best of an
approximative nature, and data from simulations appears much more
prone to noise, and systematic deviations are observed from
theoretical lines if they are continued into the unstable
phase. Qualitatively the theory is however able to capture the shape
of the octave plots, in particular their left-skewness.

\subsection{Finite size effects}
\begin{figure}[t]\vspace{3em}
\centerline{\includegraphics[width=0.35\textwidth]{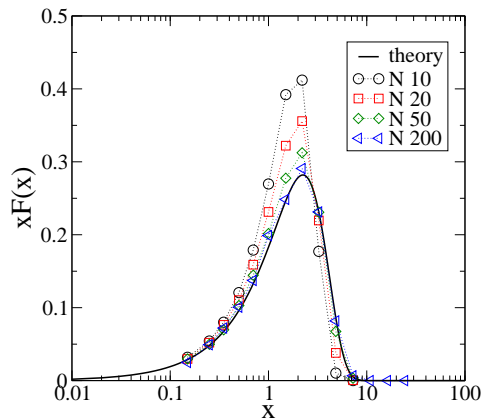}}
\caption{(Color online) $\Gamma = 0$, $u=0.4$. The line is from
 theory, valid in the thermodynamic limit $N\to\infty$. Parameters are chosen such that the system is in the stable phase, but close to the transition point of the infinite system. Markers are from simulations, $N=10$, $20$,
 $50$, $200$ respectively (averages over up to $10000$ samples are taken for small system sizes).}
\label{fig:finitesize}
																				    \end{figure}

Our theoretical analysis based on methods from statistical physics is mostly concerned with the limit of an infinite number of species in the ecosystem, $N\to\infty$. This is of course for analytical convenience only, but can be expected to be accurate also in the limit of large, but finite system size, as in real-world eco-networks. To study deviations from the exactly tractable infinite-size limit we discuss simulation results of the species abundance distribution of small systems in Fig. \ref{fig:finitesize}. One realizes that the distribution becomes more left-skewed as the system size $N$ is reduced, and that systematic deviations from the theoretical lines emerge for systems smaller than about $100$ species. For smaller $N$, the amplitude of the peak
gets larger. Note also that the largest possible concentration is limited by $N$  (due to the normalization $\sum_i x_i=N$), so that an effective upper cut-off is introduced for small systems, and the distribution is skewed to the left. In nature it is impossible to obtain data for species with an infinite concentration, so that the part of the curve at small and intermediate concentrations seems most relevant. Simulations indicate a trend toward more left-skewness at small system sizes. Unlike in other models of statistical physics at or near their phase transition points, we are unable to see fat-tailed broad species abundance distributions in the present model.

\subsection{Structure of the resulting food web}
$N$-species replicator equations can in the context of ecology be shown to be equivalent to set of $N-1$ coupled Lotka-Volterra (LV) equations \cite{Hofbauer_Sigmund_1988}. As discussed in \cite{Tokita_2004, Tokita_2006} the following transformation of variables
\BE
y_i = x_i / x_M ~~~(i=1,2,\dots,N)
\EE
\BE
r_i = w_{iM} - w_{MM} = w_{iM} + p
\EE
\BE
b_{ij} = w_{ij} - w_{Mj}
\EE
renders the replicator system studied in the previous sections equivalent to Lotka-Volterra equations of the form
\BE
\frac{dy_i}{dt}=y_i\left(r_i - \sum_j^{N-1} b_{ij} y_j\right).
\EE
The `resource species' $M\in\{1,\dots,N\}$ can here be chosen
arbitrarily, note that one then has $y_M=1$ by construction, leading
to an $N-1$ dimensional system of LV equations. The ecological
interspecies interactions $b_{ij}$ are again of a Gaussian random
form, but have different correlations than the couplings $w_{ij}$ of
the original replicator system. For $\Gamma=1$ and $\Gamma=-1$ in
particular, the $b_{ij}$ need not carry the symmetry (anti-symmetry
respectively for $\Gamma=-1$) of the couplings $w_{ij}$.  The LV model
describes an interaction network of species, where the interaction
between any given pair $(i,j)$ of species $(i\neq j$) can be of a
mutualistic type ($b_{ij}$ and $b_{ji}$ both positive), of the
competitive type ($b_{ij}$ and $b_{ji}$ both negative), or $i$ and $j$
can have a prey-predator relationship (one of the couplings positive,
the other negative). These cases are summarized in Table
\ref{tab:links}. The intraspecies interaction $b_{ii}$ is given by
$b_{ii}=w_{ii}-w_{Mi}=-p-w_{Mi}=-r_i$. $r_i$ is here the intrinsic
growth rate of species $i$ in the LV equations, and follows a Gaussian
distribution of mean $p$ and variance $1/N$. In particular, in finite
systems, $r_i$ is positive with probability
$\frac{1}{2}\left(1+\erf(\sqrt{N/2}p)\right)$. The parameter $p$($=2u$) can thus be interpreted as the `productivity' of the community (the larger $p$ the more species have positive growth rate). Note also that the average growth rate $N^{-1}\sum_i \overline{r_i}$ is given by $p$.

\begin{figure}[t]\vspace{3em}
\centerline{\includegraphics[width=0.30\textwidth]{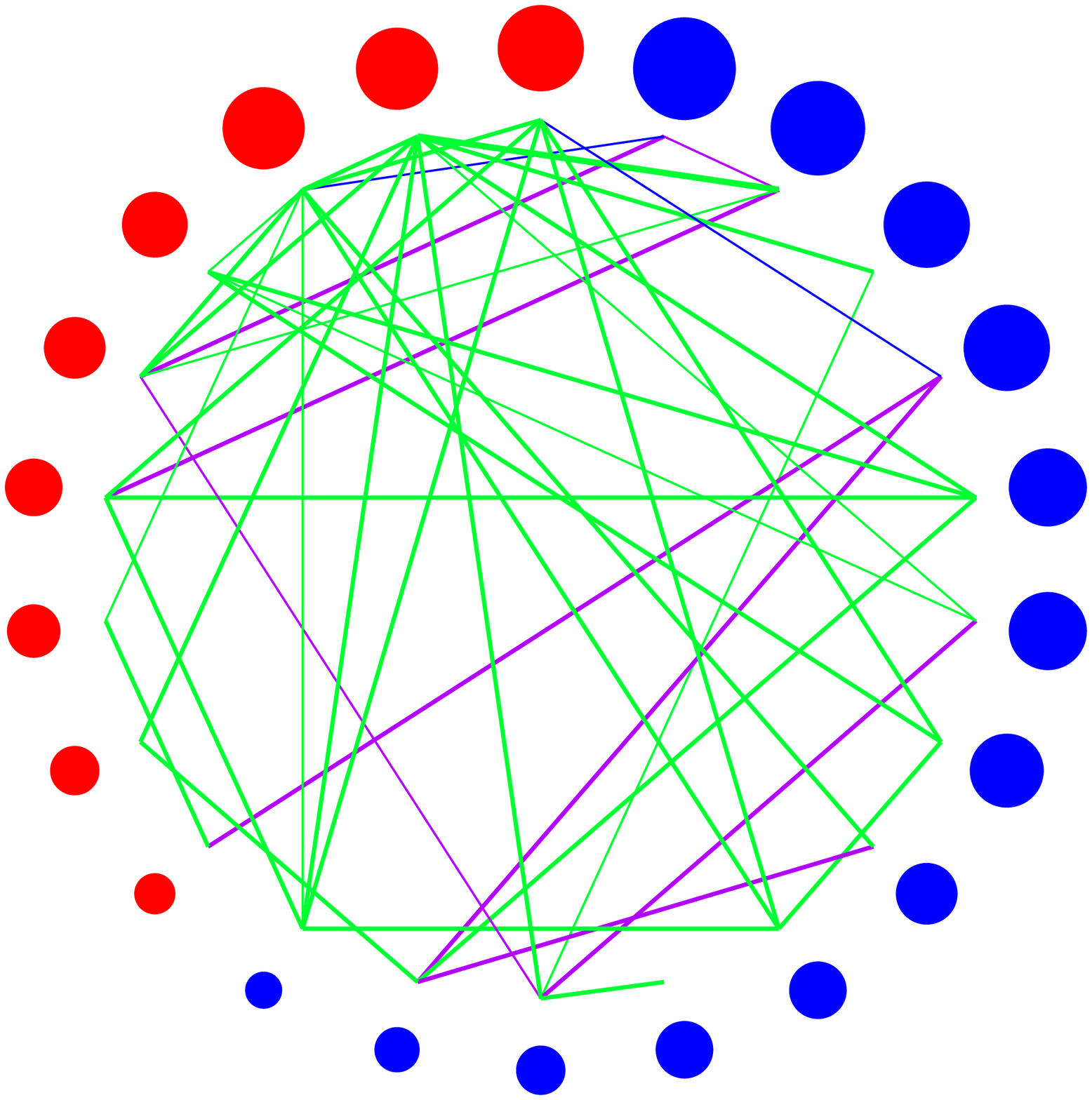}}
\caption{(Color) Network of interspecies interactions for $w=1,\Gamma=1, N=100, u=0.4$.}
\label{fig:topology_G1}
\end{figure}

\begin{figure}[t]\vspace{3em}
\centerline{\includegraphics[width=0.30\textwidth]{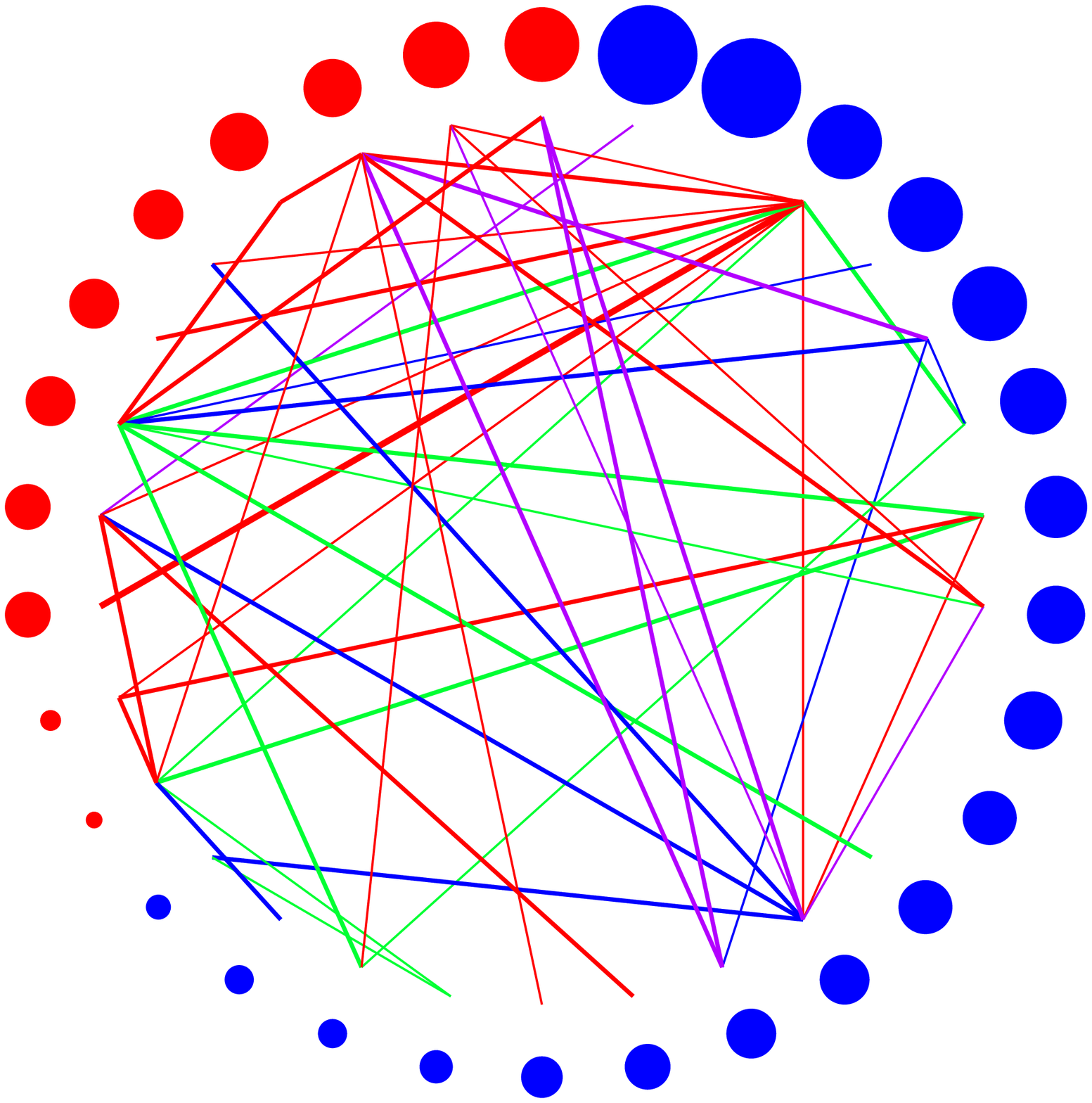}}
\caption{(Color) Network of interspecies interactions for $w=1,\Gamma=0, N=100, u=0.2$.}
\label{fig:topology_G0}
\end{figure}

\begin{figure}[t]\vspace{3em}
\centerline{\includegraphics[width=0.30\textwidth]{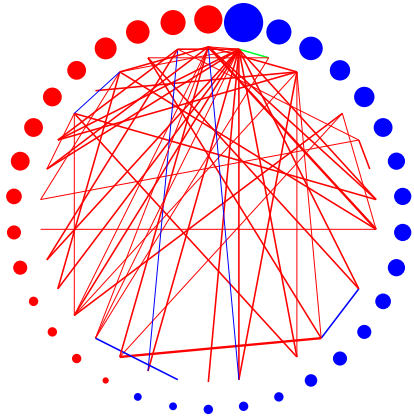}}
\caption{(Color) Network of interspecies interactions for $w=1,\Gamma=-1, N=50, u=0.2$. }
\label{fig:topology_G-1}
\end{figure}

In Figs. \ref{fig:topology_G1}, \ref{fig:topology_G0} and
\ref{fig:topology_G-1} we depict the food webs in the stationary state
of the replicator (or equivalently LV) dynamics. Disks in these
figures represent species, where species with a positive intrinsic
growth rate ($r_i>0$) are shown as blue disks, and species with
negative growth rate are depicted as red disks. Upon ordering
surviving species such that $r_1\geq r_2 \geq \dots\geq 0 \geq\dots\geq
r_S$, the radius of the disk representing species $i$ is chosen to be
proportional to $|log (r_1)|/|log |r_i||$. Note that the variance of interaction strengths scales as $1/N$ in our model, i.e.
$w_{ij}\sim O(1/\sqrt{N})$, so for small $u$ we can expect that $|r_i| < 1$ with large probability for any $i$ (we have checked that $|r_i|<1$ for all $i$ for the data shown in  Figs. \ref{fig:topology_G1}, \ref{fig:topology_G0} and
\ref{fig:topology_G-1}). Since $|log |r_i||$ is monotonically decreasing function of $|r_i|$ in the interval $0<|r_i|<1$, larger blue disks hence mean larger productivity (i.e. fast growing species if interactions $b_{ij}$ are switched off), and large red represent large anti-productivity (i.e. species with quickly decaying concentration in the absence of interactions in the LV system). Links between species are shown in the figures only if the effective
interaction exceeds a certain threshold (i.e if max($|b_{ij}|,|b_{ji}|$) $>
0.6*b_{max}$, where $b_{max}=$max($b_{ij}$) $\forall i,j$). The thickness of
each link in the Figures is in proportion to max($|b_{ij}|$, $|b_{ji}|$).

The different types of interactions (see Table \ref{tab:links}) are represented by different colors: green links denote mutualistic interactions, violet competitive interactions, blue lines denote cases where a more productive species $i$ exploits a less productive one ($j>i$, assuming species are ordered such that $r_1\geq r_2\geq ...\geq r_S$) and red the reverse case of exploitation. 

In Fig. \ref{fig:topology_G1} we depict a resulting food web for the
case of symmetric interactions ($\Gamma=1$), no red links are observed
in this case, as already reported in previous work
\cite{Tokita_2006}. On the other hand, one can see red links in
Fig. \ref{fig:topology_G0} and Fig. \ref{fig:topology_G-1}. For
$\Gamma=1$ the inter-species relationships are hence almost all
mutualistic, i.e. there are no prey-predator
type interactions in the equivalent Lotka-Volterra system. On the
other hand, for $\Gamma=-1$ the relationships are almost all of the prey-predator type and mutualistic enhancing interactions are found only very rarely in the Lotka-Volterra system. The case of uncorrelated couplings in the replicator dynamics, $\Gamma=0$, is an intermediate state.  Finally, while we show the network
topology only for small values of the co-operation pressure $u$ in the figures, we note that with larger $u$, the network becomes more dense and of a more homogeneous structure.

\begin{table}
\begin{center}

\begin{tabular}{|c||c|c|c|c|}
\hline
 Links & green & violet & blue & red\\
\hline\hline
($b_{ij}, b_{ji}$) & (+, +) & (-, -) & (+, -) & (-, +)\\
\hline
interaction & mutual & competitive & $i$ consumes $j$ & $j$ consumes $i$             \\
\hline
\end{tabular}
\end{center}
\caption{Represented the interspecies interactions as colored
 links. Species are assumed to be ordered such that $r_i>r_j$, for $i<j$.}
\label{tab:links}
            \end{table}


\section{System with heterogeneous co-operation pressure}
\begin{figure}[t]\vspace{3em}
\centerline{\includegraphics[width=0.4\textwidth]{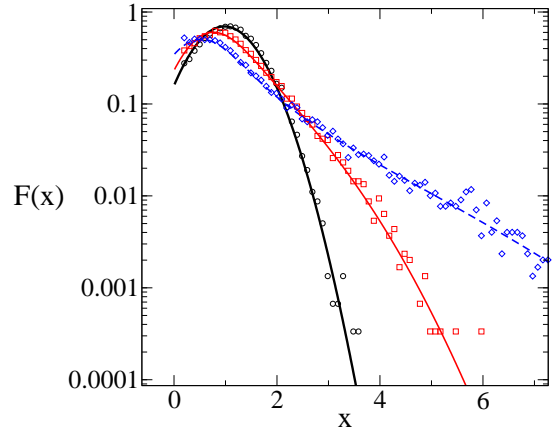}}
\caption{(Color online) Linear-log plot of distribution $F(x)$ of surviving species for a system with heterogeneous co-operation pressures drawn from a flat distribution over $[1-\sigma,1+\sigma]$ where $\sigma=0.1,0.5,0.75$ from top to bottom at the maximum. Symbols are from simulations ($\Gamma=0$, $w=1$, $N=300$ species, averages over $100$ samples), solid lines from the generating functional fixed-point theory (note that for reasons of clarity we plot $F(x)$ not $xF(x)$ in contrast with other figures of previous sections). }
\label{fig:het}
\end{figure}

Heterogeneity between species is in the present model represented by the random interactions $w_{ij}$. A second layer of diversity can be introduced, by making the co-operation pressure $u$ species dependent, i.e. to use
\be
f_i[\bx]=-2u_i x_i+\sum_{ij}w_{ij}x_j
\ee
as the fitness of species $i$, where now $u_i$ carries an explicit index $i$ and may be different from species to species. This model has been introduced and studied with generating functional techniques in \cite{Gallapreprint}. In this section we will briefly discuss how adding heterogeneity of this type effects the distribution of surviving species, and will show how it can give rise to non-Gaussian abundance distributions and how these can be computed from the statistical mechanics theory. Specifically we will draw the $\{u_i\}$ from a flat distribution over an interval $[u_0-\sigma,u_0+\sigma]$, so that $u_0$ controls the mean co-operation pressure, and $\sigma\geq 0$ is variability over the ensemble of species. The generating functional analysis is straightforward, but leads to an {\em ensemble} of effective species processes, one for each co-operation pressure present in the population. A fixed-point ansatz then leads to coupled equations for the static order parameters $Q,\chi,\Delta$, expressed as integrals over the distribution of co-operation pressures, as reported in \cite{Gallapreprint}. For $x>0$ the distribution of concentration of surviving species is then found as
\be
F(x)=\frac{1}{2\sigma}\int_{u_0-\sigma}^{u_0+\sigma}du \frac{M(u)}{\sqrt{2\pi\lambda}} \exp\left(-\frac{(
\Delta-\frac{M(u)}{\sqrt{\lambda}}x)^2}{2}
\right),
\ee
where $M(u)=2u+w^2\Gamma\chi$, $\lambda=w^2 Q$. This is a superposition of cut-off Gaussians, with varying mean and variances, and may hence for sufficient width $\sigma$ of the distribution of co-operation pressures be of non-Gaussian shape. This is indeed observed in Fig. \ref{fig:het}, where we depict $F(x)$ in a linear-log scale for various degrees of heterogeneity in the co-operation pressures. For small values of the width $\sigma$, the resulting function distribution $F(x)$ is relatively close to being Gaussian, but can develop slowly decaying tails, and non-trivial kurtosis if the co-operation pressures become sufficiently variable across species.

\section{Higher-order interaction}
\begin{figure}[t!]\vspace{3em}
\centerline{\includegraphics[width=0.4\textwidth]{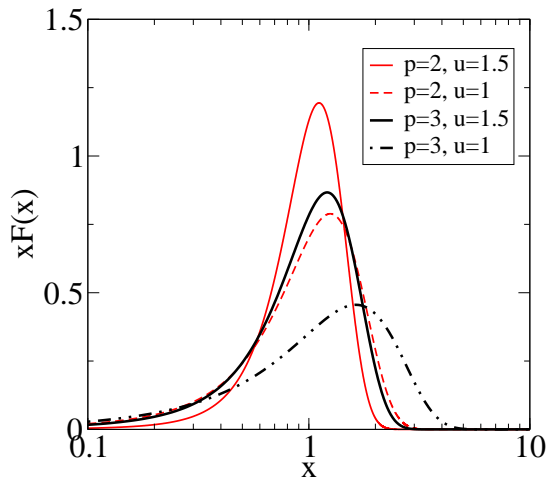}}
\caption{(Color online) Abundance distribution $xF(x)$ for the system with $2$-species and $3$-species interaction. $\Gamma=0$, the co-operation pressure is set as in indicated in the legend.}
\label{fig:p2p3_G0}
\end{figure}

Up to now we have only considered the case of pairwise interaction
between species. Generalization to higher-order interactions is possible
and has been considered for example in \cite{de_Oliveira_Fontanari_2000,
Galla_2006a}. A random community model with $p$-body interaction between  species can be defined as follows
\BE\label{eq:higher}
\frac{d}{dt}x_i(t)&=&-x_i(t)\bigg[2ux_i(t)\nonumber\\
&&+\sum_{(i_2,\dots,i_{p})\in
    M^{(p)}_i}J^{\,i}_{i_2,i_3,\dots,i_{p}}x_{i_2}(t)x_{i_3}(t)\cdots
  x_{i_{p}}(t)\nonumber\\
&&-\nu(t)\bigg], 
\EE with $p$ a fixed integer and
where $M^{(p)}_i=\{(i_2,\dots,i_{p}):1\leq i_2<i_3<\dots<i_{p}\leq N;
i_2,\dots,i_{p}\neq i\}$. The coupling tensor is again assumed to be taken from  Gaussian distribution with moments
\BE
\overline{(J^{\,i_1}_{i_2,\dots,i_p})^2}=\frac{p!}{2N^{p-1}},\nonumber \\
\overline{J^{\,i_1}_{i_2,\dots,i_p}J^{i_k}_{i_1,\dots,i_{k-1},i_{k+1},\dots,i_p}}&=&\Gamma\frac{p!}{2N^{p-1}}.
\EE
We will consider $p=3$ in the following. A generating functional and fixed-point analysis then leads to self-consistent equations
\BE
\frac{M}{\sqrt{\lambda}}&=&\int_{-\infty}^{\Delta} Dz (\Delta-z),\label{eq:firsteqp}\\
\frac{Q M^2}{\lambda}&=&\int_{-\infty}^{\Delta} Dz (\Delta-z)^2,\label{eq:secondeqp}\\
-M \chi&=&\int_{-\infty}^\Delta Dz.\label{eq:thirdeqp},
\EE
where $Dz=\frac{1}{\sqrt{2\pi}}e^{-z^2/2}dz$ again denotes the standard
Gaussian measure. These equations are very similar to the ones derived above for the case $p=2$, differences are only to be found in the detailed expressions for the quantities $M$ and $\lambda$, which now read $\lambda=\frac{3Q^2}{2}$, 
$M =2u+3 \Gamma Q \chi $. We have $\Delta= - \nu/\sqrt{\lambda}$ as before.

Results for species abundance distribution of a replicator system with
$3$-species interaction are depicted  in Fig. \ref{fig:p2p3_G0} (for
uncorrelated couplings, $\Gamma=0$), and compared to the case $p=2$ at
otherwise unchanged parameters. For reasons of clarity we do not show
results from numerical simulations, even though we have performed
numerical tests in the ergodic stable phase and find reasonable
agreement with the theoretical predictions. All other parameters kept
equal, a $3$-body interaction appears to shift the peak of the
distribution to the right, and to reduce its height, while increasing
its width and left-skewness. Our findings thus suggest that higher-order interactions may add to the diversity of the ecological community, i.e. increase the variance of species concentrations at stationarity.

\section{Summary and concluding remarks}
In summary we have presented a detailed discussion of species abundance
relations resulting from the evolutionary dynamics of random replicator
systems. Based on dynamical techniques of statistical mechanics of
disordered systems we have extended the work of
\cite{Tokita_2004,Tokita_2006} to the case of asymmetric and
anti-symmetric coupling matrices, and have also taken into account
higher-order interaction modes and systems in which species are
subject to heterogeneous co-operation pressures. These systems
typically show a phase transition between a stable, ergodic regime and
an unstable phase, in which the final state of the system depends on
initial conditions. Based on a fixed-point ansatz the statistical
mechanics theory is able to deliver exact analytical predictions for
the resulting species abundance relations in the limit of infinite
system-size, and computer simulations of the replicator dynamics are
in perfect agreement with theoretical predictions. The key findings of
our analysis are the following: (i) with larger co-operation pressure,
regardless of inter-species interaction, the diversity of the
population increases, (ii) we derive species-poor and species-rich RAR
for symmetric interaction and species-rich RAR for asymmetric
interaction, (iii) we find that the abundance distributions are
typically similar to a lognormal distribution, and of a left-skewed
type in our model, not too dissimilar from empirical data, (iv)
visualizing the food-web structure of surviving species, and
distinguishing between different types of pairwise species
interactions gives insight into the stable relationship between
species at stationarity, in particular symmetric interactions favor
mutualistic relations, whereas anti-symmetric couplings tend to lead
to one-sided exploitation of some species by others, (v) survival
functions of systems with heterogeneous co-operation pressure can
display highly non-Gaussian survival functions with long tails, (vi)
in finite systems our theory is not applicable, and systematic
deviations are observed. In contrast with other disordered systems SAD
are not found to be fat-tailed or skewed to the right near the
transition of the infinite-size model.

The techniques we employ to study species abundance in random replicator systems are in the present context limited to fully connected random communities with Gaussian interactions. Extension to more realistic distributions of couplings may here be of interest, and similarly more realistic food-web topologies (see e.g. \cite{Dorogovtsev_Mendes_2003} or \cite{Dunne_etal_2002a} and references therein) could be taken into account in future work. Methods from disordered systems theory can be adapted to those cases as well, and further studies would most likely be based on cavity methods or other tools used for finite-connectivity disordered systems \cite{Hartmann_Weigt_2005, Coolen_et_al_2005}.

There is currently also much interest in the relationship between deterministic models of population dynamics (defined through rate equations, e.g. the above replicator dynamics) and stochastic individual-based models \cite{McKane_Newman_2004,McKane_Newman_2005}.  It has here been seen that demographic stochasticity in models with a finite-number of individuals can induce behavior quite different from models based on rate equations It may hence be of interest to investigate finite microscopic individual-based analogues of random replicator systems (for example based on Moran dynamics) and to compare their dynamical behavior to that of the mean-field replicator system. Individual-based versions of systems with randomly drawn reaction rates have to our knowledge not been considered in the literature. This is indeed an interesting line of potential future work, although caution is appropriate when it comes to analytical approaches, as the randomness of interactions may make closed-form solutions of such models very difficult.

 It is hoped that our work may serve as a starting point for future
 studies in these directions, and that analysis of random community
 models of theoretical ecology based on methods from statistical
 mechanics may hence contribute to an understand of issues related to
 the diversity-stability debate as mentioned in the introduction.

\begin{acknowledgements} 
This work was supported by EU NEST No. 516446 COMPLEXMARKETS, by IST
STREP GENNETEC, contract number 034952. TG acknowledges support through 
an RCUK Fellowship (RCUK reference EP/E500048/1).  YY and KT are
partially supported by The 21st Century COE program `Towards a new
basic science: depth and synthesis'. KT acknowledges support by
grants-in-aid from MEXT, Japan (No. 14740232 and 17540383) and through
the priority area `Systems Genomics'.
\end{acknowledgements}

\section*{Appendix}
The analysis of disordered systems by means of generating functional
is a useful and powerful method, especially because it does not require the
existence of a Lyapunov function, and is hence not limited to systems
with symmetric interaction matrices. In this appendix we briefly
outline the main mathematical steps and concepts of this technique.
Further details can be found in a broad spectrum of sources in the
literature \cite{DeDominicis_1978, Mezard_etal_1987, Coolen_2000,
Coolen_et_al_2005}.

The basic idea is to reduce a
high-dimensional system with random couplings to an effective process
for a respresentative (mean-field) particle. These processes are
typically non-Markovian, even if the original system is Markovian, and
subject to colored noise. If
$\bx(t)=(x_1(t),\dots,x_N(t))$ represents a trajectory of the
microscopic system (subject to random interactions), then the starting point
of the analysis is the dynamical partition function (or generating
functional)
\be\label{eq:gfZ} Z[\psi]=\avg{\exp\left[\sum_t
i\psi(t)x_i(t)\right]},
\ee
where $\avg{\cdots}$ represents an average over {\em all} possible
trajectories of the system. The dynamic partition function can hence
be expressed as a path-integral over all such trajectories, and
written in the form
\be
Z[\psi]=\int D\bx~ \delta(\mbox{eq. of motion})~e^{i\sum_t
\psi(t)x_i(t)}.
\ee
By `equations of motion' we here mean the microscopic equations
governing the dynamics, in our case the replicator equations
Eq. (\ref{eq:repl}), they contain the quenched disorder of the problem
(i.e. the random couplings). The analysis proceeds by writing the
delta-functions in their Fourier representation by means of conjugate
variables $\{\widehat
x_i(t)\}$, subsequently performing the average over the disorder, and
then by introducing suitable macroscopic order parameters, such as
e.g. the correlation function $C(t,t')=N^{-1}\sum_i x_i(t)x_i(t')$ and
the response function $G(t,t')=iN^{-1}\sum_i x_i(t)\widehat x_i(t)$.
In the thermodynamic limit, $N\to\infty$, an effective theory for $C$
and $G$ is then derived, expressed as a self-consistent problem
involving the above mentioned effective single-particle process in
conjunction with self-consistent relations for correlation and
response functions. As seen in Eqs. (\ref{eq:effproc},\ref{eq:effproc_eta}) the
effective process makes reference to $C$ and $G$, and on the other
hand these order parameters are to be computed self-consistently as
averages over the ensemble of effective-particle trajectories
(\ref{eq:effproc_CG}).

For general systems the effective single-particle problem can be
addressed by suitable numerical schemes \cite{Eissfeller_Opper_1992}. In the
case of the replicator problem further analytical progress is possible
based on the observation that the system attains a fixed point at
sufficiently large co-operation pressure \cite{Opper_Diederich_1992}. In
this regime trajectories become effectively time-independent
asymptotically, and further simplification is possible yielding Eqs.
(\ref{eq:firsteq}-\ref{eq:thirdeq}). Details of these steps can be
found in \cite{Opper_Diederich_1992} and \cite{Galla_2006a}.
\bibliographystyle{apsrev}

\begin{thebibliography}{53}
\expandafter\ifx\csname natexlab\endcsname\relax\def\natexlab#1{#1}\fi
\expandafter\ifx\csname bibnamefont\endcsname\relax
  \def\bibnamefont#1{#1}\fi
\expandafter\ifx\csname bibfnamefont\endcsname\relax
  \def\bibfnamefont#1{#1}\fi
\expandafter\ifx\csname citenamefont\endcsname\relax
  \def\citenamefont#1{#1}\fi
\expandafter\ifx\csname url\endcsname\relax
  \def\url#1{\texttt{#1}}\fi
\expandafter\ifx\csname urlprefix\endcsname\relax\def\urlprefix{URL }\fi
\providecommand{\bibinfo}[2]{#2}
\providecommand{\eprint}[2][]{\url{#2}}

\bibitem[{\citenamefont{McCann}(2000)}]{McCann_2000}
\bibinfo{author}{\bibfnamefont{K.~S.} \bibnamefont{McCann}},
  \bibinfo{journal}{Nature} \textbf{\bibinfo{volume}{405}},
  \bibinfo{pages}{228} (\bibinfo{year}{2000}).

\bibitem[{\citenamefont{MacArthur}(1955)}]{MacArthur_1955}
\bibinfo{author}{\bibfnamefont{R.~H.} \bibnamefont{MacArthur}},
  \bibinfo{journal}{Ecology} \textbf{\bibinfo{volume}{36}},
  \bibinfo{pages}{533} (\bibinfo{year}{1955}).

\bibitem[{\citenamefont{Elton}(1958)}]{Elton_1958}
\bibinfo{author}{\bibfnamefont{C.~S.} \bibnamefont{Elton}},
  \emph{\bibinfo{title}{The ecology of invasion by animals and plants}}
  (\bibinfo{publisher}{Mathuen}, \bibinfo{address}{London},
  \bibinfo{year}{1958}).

\bibitem[{\citenamefont{Gardner and Ashby}(1970)}]{Gardner_Ashby_1970}
\bibinfo{author}{\bibfnamefont{M.~R.} \bibnamefont{Gardner}} \bibnamefont{and}
  \bibinfo{author}{\bibfnamefont{W.~R.} \bibnamefont{Ashby}},
  \bibinfo{journal}{Nature} \textbf{\bibinfo{volume}{228}},
  \bibinfo{pages}{784} (\bibinfo{year}{1970}).

\bibitem[{\citenamefont{May}(1972)}]{May_1972}
\bibinfo{author}{\bibfnamefont{R.~M.} \bibnamefont{May}},
  \bibinfo{journal}{Nature} \textbf{\bibinfo{volume}{238}},
  \bibinfo{pages}{413} (\bibinfo{year}{1972}).

\bibitem[{\citenamefont{May}(1974)}]{May_1974}
\bibinfo{author}{\bibfnamefont{R.~M.} \bibnamefont{May}},
  \emph{\bibinfo{title}{Stability and complexity in model ecosystems, 2nd ed.}}
  (\bibinfo{publisher}{Princeton Univ. Press}, \bibinfo{address}{Princeton},
  \bibinfo{year}{1974}).

\bibitem[{\citenamefont{Pimm}(1991)}]{Pimm_1991}
\bibinfo{author}{\bibfnamefont{S.~L.} \bibnamefont{Pimm}},
  \emph{\bibinfo{title}{{T}he balance of nature?}} (\bibinfo{publisher}{Chicago
  University Press}, \bibinfo{address}{Chicago}, \bibinfo{year}{1991}).

\bibitem[{\citenamefont{Rozdilsky and Stone}(2001)}]{Rozdilsky_Stone_2001}
\bibinfo{author}{\bibfnamefont{I.~D.} \bibnamefont{Rozdilsky}}
  \bibnamefont{and} \bibinfo{author}{\bibfnamefont{L.~S.} \bibnamefont{Stone}},
  \bibinfo{journal}{Ecol. Lett.} \textbf{\bibinfo{volume}{4}},
  \bibinfo{pages}{397} (\bibinfo{year}{2001}).

\bibitem[{\citenamefont{Chawanya and Tokita}(2002)}]{Chawanya_Tokita_2002}
\bibinfo{author}{\bibfnamefont{T.}~\bibnamefont{Chawanya}} \bibnamefont{and}
  \bibinfo{author}{\bibfnamefont{K.}~\bibnamefont{Tokita}},
  \bibinfo{journal}{J. Phys. Soc. Jpn.} \textbf{\bibinfo{volume}{71}},
  \bibinfo{pages}{429} (\bibinfo{year}{2002}).

\bibitem[{\citenamefont{Lawler and Morin}(1993)}]{Lawler_Morin_1993}
\bibinfo{author}{\bibfnamefont{S.~P.} \bibnamefont{Lawler}} \bibnamefont{and}
  \bibinfo{author}{\bibfnamefont{P.~J.} \bibnamefont{Morin}},
  \bibinfo{journal}{Am. Nat.} \textbf{\bibinfo{volume}{141}},
  \bibinfo{pages}{675} (\bibinfo{year}{1993}).

\bibitem[{\citenamefont{McCann et~al.}(1998)\citenamefont{McCann, Hastings, and
  Huxel}}]{McCann_Hastings_Huxel_1998}
\bibinfo{author}{\bibfnamefont{K.}~\bibnamefont{McCann}},
  \bibinfo{author}{\bibfnamefont{A.}~\bibnamefont{Hastings}}, \bibnamefont{and}
  \bibinfo{author}{\bibfnamefont{G.~R.} \bibnamefont{Huxel}},
  \bibinfo{journal}{Nature} \textbf{\bibinfo{volume}{395}},
  \bibinfo{pages}{794} (\bibinfo{year}{1998}).

\bibitem[{\citenamefont{Neutel et~al.}(2002)\citenamefont{Neutel, Heesterbeek,
  and de~Ruiter}}]{Neutel_etal_2002}
\bibinfo{author}{\bibfnamefont{A.-M.} \bibnamefont{Neutel}},
  \bibinfo{author}{\bibfnamefont{J.~A.~P.} \bibnamefont{Heesterbeek}},
  \bibnamefont{and} \bibinfo{author}{\bibfnamefont{P.~C.}
  \bibnamefont{de~Ruiter}}, \bibinfo{journal}{Science}
  \textbf{\bibinfo{volume}{296}}, \bibinfo{pages}{1120} (\bibinfo{year}{2002}).

\bibitem[{\citenamefont{Brown}(1995)}]{Brown_1995}
\bibinfo{author}{\bibfnamefont{J.~H.} \bibnamefont{Brown}},
  \emph{\bibinfo{title}{Macroecology}} (\bibinfo{publisher}{University of
  Chicago Press}, \bibinfo{year}{1995}).

\bibitem[{\citenamefont{Rosenzweig}(1995)}]{Rosenzweig_1995}
\bibinfo{author}{\bibfnamefont{M.~L.} \bibnamefont{Rosenzweig}},
  \emph{\bibinfo{title}{{S}pecies {D}iversity in {S}pace and {T}ime}}
  (\bibinfo{publisher}{Cambridge Univ. Press}, \bibinfo{address}{Cambridge},
  \bibinfo{year}{1995}).

\bibitem[{\citenamefont{May}(1975)}]{May_1975}
\bibinfo{author}{\bibfnamefont{R.~M.} \bibnamefont{May}}, \bibinfo{journal}{In
  {\it Ecology and Evolution of Communities}, Ed. M. L. Cody and J. M. Diamond,
  Belknap} pp. \bibinfo{pages}{81--120} (\bibinfo{year}{1975}).

\bibitem[{\citenamefont{Sugihara}(1980)}]{Sugihara_1980}
\bibinfo{author}{\bibfnamefont{G.}~\bibnamefont{Sugihara}},
  \bibinfo{journal}{Am. Nat.} \textbf{\bibinfo{volume}{116}},
  \bibinfo{pages}{770} (\bibinfo{year}{1980}).

\bibitem[{\citenamefont{Nee et~al.}(1991)\citenamefont{Nee, Harvey, and
  May}}]{Nee_Harvey_May_1991}
\bibinfo{author}{\bibfnamefont{S.}~\bibnamefont{Nee}},
  \bibinfo{author}{\bibfnamefont{P.~H.} \bibnamefont{Harvey}},
  \bibnamefont{and} \bibinfo{author}{\bibfnamefont{R.~M.} \bibnamefont{May}},
  \bibinfo{journal}{Proc. R. Soc. Lond. B} \textbf{\bibinfo{volume}{243}},
  \bibinfo{pages}{161} (\bibinfo{year}{1991}).

\bibitem[{\citenamefont{Tokeshi}(1998)}]{Tokeshi_1998}
\bibinfo{author}{\bibfnamefont{M.}~\bibnamefont{Tokeshi}},
  \emph{\bibinfo{title}{{S}pecies {C}oexistence}}
  (\bibinfo{publisher}{Blackwell}, \bibinfo{year}{1998}).
\bibitem[{\citenamefont{Hall et~al.}(2002)\citenamefont{Hall, Christensen, {di
  Collobiano}, and Jensen}}]{Hall_etal_2002}

\bibinfo{author}{\bibfnamefont{M.}~\bibnamefont{Hall}},
  \bibinfo{author}{\bibfnamefont{K.}~\bibnamefont{Christensen}},
  \bibinfo{author}{\bibfnamefont{S.~A.} \bibnamefont{{di Collobiano}}},
  \bibnamefont{and} \bibinfo{author}{\bibfnamefont{H.~J.}
  \bibnamefont{Jensen}}, \bibinfo{journal}{Phys. Rev. E}
  \textbf{\bibinfo{volume}{66}}, \bibinfo{pages}{011904}
  (\bibinfo{year}{2002}).

\bibitem[{\citenamefont{Hubbell}(2001)}]{Hubbell_2001}
\bibinfo{author}{\bibfnamefont{S.~P.} \bibnamefont{Hubbell}},
  \emph{\bibinfo{title}{{T}he {U}nified {N}eutral {T}heory of {B}iodiversity
  and {B}iogeography}} (\bibinfo{publisher}{Princeton University Press},
  \bibinfo{address}{Princeton}, \bibinfo{year}{2001}).

\bibitem[{\citenamefont{Volkov et~al.}(2003)\citenamefont{Volkov, Banavar,
  Hubbel., and Maritan}}]{Volkov_etal_2003}
\bibinfo{author}{\bibfnamefont{I.}~\bibnamefont{Volkov}},
  \bibinfo{author}{\bibfnamefont{J.~R.} \bibnamefont{Banavar}},
  \bibinfo{author}{\bibfnamefont{S.~P.} \bibnamefont{Hubbel.}},
  \bibnamefont{and} \bibinfo{author}{\bibfnamefont{A.}~\bibnamefont{Maritan}},
  \bibinfo{journal}{Nature} \textbf{\bibinfo{volume}{424}},
  \bibinfo{pages}{1035} (\bibinfo{year}{2003}).

\bibitem[{\citenamefont{Etienne and Olff}(2004)}]{Etienne_Olff_2004}
\bibinfo{author}{\bibfnamefont{R.~S.}~\bibnamefont{Etienne}} \bibnamefont{and}
  \bibinfo{author}{\bibfnamefont{H.}~\bibnamefont{Olff}},
  \bibinfo{journal}{Ecol. Lett.} \textbf{\bibinfo{volume}{7}},
  \bibinfo{pages}{170} (\bibinfo{year}{2004}).

\bibitem[{\citenamefont{Alonso and McKane}(2004)}]{Alonso_McKane_2004}
\bibinfo{author}{\bibfnamefont{D.}~\bibnamefont{Alonso}} \bibnamefont{and}
  \bibinfo{author}{\bibfnamefont{A.~J.} \bibnamefont{McKane}},
  \bibinfo{journal}{Ecol. Lett.} \textbf{\bibinfo{volume}{7}},
  \bibinfo{pages}{901} (\bibinfo{year}{2004}).

\bibitem[{\citenamefont{Etienne}(2005)}]{Etienne_2005}
\bibinfo{author}{\bibfnamefont{R.~S.}~\bibnamefont{Etienne}},
  \bibinfo{journal}{Ecol. Lett.} \textbf{\bibinfo{volume}{8}},
  \bibinfo{pages}{253} (\bibinfo{year}{2005}).

\bibitem[{\citenamefont{Alonso et~al.}(2006)\citenamefont{Alonso, Etienne, and McKane}}]{Alonso_etal_2006}
\bibinfo{author}{\bibfnamefont{D.}~\bibnamefont{Alonso}},
  \bibinfo{author}{\bibfnamefont{R.~S.} \bibnamefont{Etienne}},
  \bibnamefont{and} \bibinfo{author}{\bibfnamefont{A.~J.}~\bibnamefont{McKane}},
  \bibinfo{journal}{Trends in Ecol. Evol.} \textbf{\bibinfo{volume}{21}},
  \bibinfo{pages}{451} (\bibinfo{year}{2006}).

\bibitem[{\citenamefont{Etienne and Alonso}(2007)}]{Etienne_Alonso_2007}
\bibinfo{author}{\bibfnamefont{R.~S.}~\bibnamefont{Etienne}} \bibnamefont{and}
  \bibinfo{author}{\bibfnamefont{D.}~\bibnamefont{Alonso}},
  \bibinfo{journal}{J. Stat. Phys.} \textbf{\bibinfo{volume}{128}},
  \bibinfo{pages}{485} (\bibinfo{year}{2007}).

\bibitem[{\citenamefont{Diederich and Opper}(1989)}]{Diederich_Opper_1989}
\bibinfo{author}{\bibfnamefont{S.}~\bibnamefont{Diederich}} \bibnamefont{and}
  \bibinfo{author}{\bibfnamefont{M.}~\bibnamefont{Opper}},
  \bibinfo{journal}{Phys. Rev. A} \textbf{\bibinfo{volume}{39}},
  \bibinfo{pages}{4333} (\bibinfo{year}{1989}).

\bibitem[{\citenamefont{Opper and Diederich}(1992)}]{Opper_Diederich_1992}
\bibinfo{author}{\bibfnamefont{M.}~\bibnamefont{Opper}} \bibnamefont{and}
  \bibinfo{author}{\bibfnamefont{S.}~\bibnamefont{Diederich}},
  \bibinfo{journal}{Phys. Rev. Lett.} \textbf{\bibinfo{volume}{69}},
  \bibinfo{pages}{1616} (\bibinfo{year}{1992}).

\bibitem[{\citenamefont{Tokita}(2004)}]{Tokita_2004}
\bibinfo{author}{\bibfnamefont{K.}~\bibnamefont{Tokita}},
  \bibinfo{journal}{Phys. Rev. Lett.} \textbf{\bibinfo{volume}{93}},
  \bibinfo{pages}{178102} (\bibinfo{year}{2004}).

\bibitem[{\citenamefont{McKane et~al.}(2000)\citenamefont{McKane, Alonso, and
  Sol\'{e}}}]{McKane_Alonso_Sole_2000}
\bibinfo{author}{\bibfnamefont{A.}~\bibnamefont{McKane}},
  \bibinfo{author}{\bibfnamefont{D.}~\bibnamefont{Alonso}}, \bibnamefont{and}
  \bibinfo{author}{\bibfnamefont{R.~V.} \bibnamefont{Sol\'{e}}},
  \bibinfo{journal}{Phys. Rev. E} \textbf{\bibinfo{volume}{62}},
  \bibinfo{pages}{8466} (\bibinfo{year}{2000}).

\bibitem[{\citenamefont{Sol\'{e} et~al.}(2002)\citenamefont{Sol\'{e}, Alonso,
  and McKane}}]{Sole_Alonso_McKane_2002}
\bibinfo{author}{\bibfnamefont{R.~V.} \bibnamefont{Sol\'{e}}},
  \bibinfo{author}{\bibfnamefont{D.}~\bibnamefont{Alonso}}, \bibnamefont{and}
  \bibinfo{author}{\bibfnamefont{A.}~\bibnamefont{McKane}},
  \bibinfo{journal}{Phil. Trans. R. Soc. B} \textbf{\bibinfo{volume}{357}},
  \bibinfo{pages}{667} (\bibinfo{year}{2002}).

\bibitem[{\citenamefont{Mazenko}(2006)}]{Mazenko_2006}
\bibinfo{author}{\bibfnamefont{G.~F.}~\bibnamefont{Mazenko}},
  \emph{\bibinfo{title}{{N}onequilibrium {S}tatistical {M}echanics}} (\bibinfo{publisher}{Wiley-VCH},
  \bibinfo{address}{Weinheim}, \bibinfo{year}{2006}).

\bibitem[{\citenamefont{Pathria}(1996)}]{Pathria_1996}
\bibinfo{author}{\bibfnamefont{R.~K.}~\bibnamefont{Pathria}},
  \emph{\bibinfo{title}{{S}tatistical {M}echanics}} (\bibinfo{publisher}{Butterworth-Heinemann},
  \bibinfo{address}{Oxford}, \bibinfo{year}{1996}).

\bibitem[{\citenamefont{Sherrington and
  Kirkpatrick}(1975)}]{Sherrington_Kirkpatrick_1978}
\bibinfo{author}{\bibfnamefont{D.}~\bibnamefont{Sherrington}} \bibnamefont{and}
  \bibinfo{author}{\bibfnamefont{S.}~\bibnamefont{Kirkpatrick}},
  \bibinfo{journal}{Phys. Rev. Lett.} \textbf{\bibinfo{volume}{35}},
  \bibinfo{pages}{1792} (\bibinfo{year}{1975}).

\bibitem[{\citenamefont{Mezard et~al.}(1987)\citenamefont{Mezard, Parisi, and
  Virasoro}}]{Mezard_etal_1987}
\bibinfo{author}{\bibfnamefont{M.}~\bibnamefont{Mezard}},
  \bibinfo{author}{\bibfnamefont{G.}~\bibnamefont{Parisi}}, \bibnamefont{and}
  \bibinfo{author}{\bibfnamefont{A.}~\bibnamefont{Virasoro}},
  \emph{\bibinfo{title}{{S}pin {G}lass {T}heory and {B}eyond}}
  (\bibinfo{publisher}{World Scientific}, \bibinfo{address}{Singapore},
  \bibinfo{year}{1987}).

\bibitem[{\citenamefont{D{\"{u}}ring et~al.}(1998)\citenamefont{D{\"{u}}ring,
  Coolen, and Sherrington}}]{During_etal_1998}
\bibinfo{author}{\bibfnamefont{A.}~\bibnamefont{D{\"{u}}ring}},
  \bibinfo{author}{\bibfnamefont{A.~C.~C.} \bibnamefont{Coolen}},
  \bibnamefont{and}
  \bibinfo{author}{\bibfnamefont{D.}~\bibnamefont{Sherrington}},
  \bibinfo{journal}{J. Phys. A: Math. Gen.} \textbf{\bibinfo{volume}{31}},
  \bibinfo{pages}{8607} (\bibinfo{year}{1998}).

\bibitem[{\citenamefont{Mimura et~al.}(2004)\citenamefont{Mimura, Kawamura, and
  Okada}}]{Mimura_2004}
\bibinfo{author}{\bibfnamefont{K.}~\bibnamefont{Mimura}},
  \bibinfo{author}{\bibfnamefont{M.}~\bibnamefont{Kawamura}}, \bibnamefont{and}
  \bibinfo{author}{\bibfnamefont{M.}~\bibnamefont{Okada}}, \bibinfo{journal}{J.
  Phys. A: Math. Gen.} \textbf{\bibinfo{volume}{37}}, \bibinfo{pages}{6437}
  (\bibinfo{year}{2004}).

\bibitem[{\citenamefont{Hofbauer and Sigmund}(1988)}]{Hofbauer_Sigmund_1988}
\bibinfo{author}{\bibfnamefont{J.}~\bibnamefont{Hofbauer}} \bibnamefont{and}
  \bibinfo{author}{\bibfnamefont{K.}~\bibnamefont{Sigmund}},
  \emph{\bibinfo{title}{{T}he {T}heory of {E}volution and {D}ynamical
  {S}ystems}} (\bibinfo{publisher}{Cambridge University Press},
  \bibinfo{address}{Cambridge}, \bibinfo{year}{1988}).

\bibitem[{\citenamefont{{de Oliveira} and
  Fontanari}(2000)}]{de_Oliveira_Fontanari_2000}
\bibinfo{author}{\bibfnamefont{V.~M.} \bibnamefont{{de Oliveira}}}
  \bibnamefont{and} \bibinfo{author}{\bibfnamefont{J.~F.}
  \bibnamefont{Fontanari}}, \bibinfo{journal}{Phys. Rev. Lett.}
  \textbf{\bibinfo{volume}{85}}, \bibinfo{pages}{4984} (\bibinfo{year}{2000}).

\bibitem[{\citenamefont{Galla}(2006)}]{Galla_2006a}
\bibinfo{author}{\bibfnamefont{T.}~\bibnamefont{Galla}}, \bibinfo{journal}{J.
  Phys. A: Math. Gen.} \textbf{\bibinfo{volume}{39}}, \bibinfo{pages}{3853}
  (\bibinfo{year}{2006}).

\bibitem[{\citenamefont{Peschel and Mende}(1986)}]{Paschel_Mende_1986}
\bibinfo{author}{\bibfnamefont{M.}~\bibnamefont{Peschel}} \bibnamefont{and}
  \bibinfo{author}{\bibfnamefont{W.}~\bibnamefont{Mende}},
  \emph{\bibinfo{title}{{T}he {P}rey-{P}redator {M}odel}}
  (\bibinfo{publisher}{SpringerVerlag}, \bibinfo{address}{Vienna},
  \bibinfo{year}{1986}).

\bibitem[{\citenamefont{Opper and Diederich}(1999)}]{Opper_Diederich_1999}
\bibinfo{author}{\bibfnamefont{M.}~\bibnamefont{Opper}} \bibnamefont{and}
  \bibinfo{author}{\bibfnamefont{S.}~\bibnamefont{Diederich}},
  \bibinfo{journal}{Comp. Phys. Commn.} \textbf{\bibinfo{volume}{{121-122}}},
  \bibinfo{pages}{141} (\bibinfo{year}{1999}).

\bibitem[{\citenamefont{{De Dominicis}}(1978)}]{DeDominicis_1978}
\bibinfo{author}{\bibfnamefont{C.}~\bibnamefont{{De Dominicis}}},
  \bibinfo{journal}{Phys. Rev. B} \textbf{\bibinfo{volume}{18}},
  \bibinfo{pages}{4913} (\bibinfo{year}{1978}).

\bibitem[{\citenamefont{Rieger}(1989)}]{Rieger_1989}
\bibinfo{author}{\bibfnamefont{H.}~\bibnamefont{Rieger}}, \bibinfo{journal}{J.
  Phys. A} \textbf{\bibinfo{volume}{22}}, \bibinfo{pages}{3447}
  (\bibinfo{year}{1989}).

\bibitem[{\citenamefont{Galla}(2008)}]{Galla_forthcoming}
\bibinfo{author}{\bibfnamefont{T.}~\bibnamefont{Galla}}, \bibinfo{journal}{(to
  be submitted)}  (\bibinfo{year}{2008}).

\bibitem[{\citenamefont{Biscari and Parisi}(1995)}]{Biscari_Parisi_1995}
\bibinfo{author}{\bibfnamefont{P.}~\bibnamefont{Biscari}} \bibnamefont{and}
  \bibinfo{author}{\bibfnamefont{G.}~\bibnamefont{Parisi}},
  \bibinfo{journal}{J. Phys. A: Math. Gen.} \textbf{\bibinfo{volume}{28}},
  \bibinfo{pages}{4697} (\bibinfo{year}{1995}).

\bibitem[{\citenamefont{Whittaker}(1970)}]{Whittaker_1970}
\bibinfo{author}{\bibfnamefont{R.~H.} \bibnamefont{Whittaker}},
  \emph{\bibinfo{title}{{C}ommunities and {E}cosystems}}
  (\bibinfo{publisher}{Macmillan}, \bibinfo{address}{New York},
  \bibinfo{year}{1970}).

\bibitem[{\citenamefont{Bazzaz}(1975)}]{Bazzaz_1975}
\bibinfo{author}{\bibfnamefont{F.~A.} \bibnamefont{Bazzaz}},
  \bibinfo{journal}{Ecology} \textbf{\bibinfo{volume}{56}},
  \bibinfo{pages}{485} (\bibinfo{year}{1975}).

\bibitem[{\citenamefont{Tokita and Yasutomi}(2003)}]{Tokita_Yasutomi_2003}
\bibinfo{author}{\bibfnamefont{K.}~\bibnamefont{Tokita}} \bibnamefont{and}
  \bibinfo{author}{\bibfnamefont{A.}~\bibnamefont{Yasutomi}},
  \bibinfo{journal}{Theor. Popul. Biol.} \textbf{\bibinfo{volume}{63}},
  \bibinfo{pages}{131} (\bibinfo{year}{2003}).

\bibitem[{\citenamefont{Fisher et~al.}(1943)\citenamefont{Fisher, Corbet, and
  Williams}}]{Fisher_Corbet_Williams_1943}
\bibinfo{author}{\bibfnamefont{R.~A.} \bibnamefont{Fisher}},
  \bibinfo{author}{\bibfnamefont{A.~S.} \bibnamefont{Corbet}},
  \bibnamefont{and} \bibinfo{author}{\bibfnamefont{C.~B.}
  \bibnamefont{Williams}}, \bibinfo{journal}{J. Anim. Ecol.}
  \textbf{\bibinfo{volume}{12}}, \bibinfo{pages}{42} (\bibinfo{year}{1943}).

\bibitem[{\citenamefont{Preston}(1962{\natexlab{a}})}]{Preston_1962a}
\bibinfo{author}{\bibfnamefont{F.~W.} \bibnamefont{Preston}},
  \bibinfo{journal}{Ecology} \textbf{\bibinfo{volume}{43}},
  \bibinfo{pages}{185} (\bibinfo{year}{1962}{\natexlab{a}}).

\bibitem[{\citenamefont{Preston}(1962{\natexlab{b}})}]{Preston_1962b}
\bibinfo{author}{\bibfnamefont{F.~W.} \bibnamefont{Preston}},
  \bibinfo{journal}{Ecology} \textbf{\bibinfo{volume}{43}},
  \bibinfo{pages}{410} (\bibinfo{year}{1962}{\natexlab{b}}).

\bibitem[{\citenamefont{MacArthur}(1957)}]{MacArthur_1957}
\bibinfo{author}{\bibfnamefont{R.~H.} \bibnamefont{MacArthur}},
  \bibinfo{journal}{Proc. Nat. Acad. Sci. USA} \textbf{\bibinfo{volume}{43}},
  \bibinfo{pages}{293} (\bibinfo{year}{1957}).

\bibitem[{\citenamefont{Tokita}(2006)}]{Tokita_2006}
\bibinfo{author}{\bibfnamefont{K.}~\bibnamefont{Tokita}},
  \bibinfo{journal}{Ecological Informatics} \textbf{\bibinfo{volume}{1}},
  \bibinfo{pages}{315} (\bibinfo{year}{2006}).

\bibitem[{\citenamefont{Galla}(2007)}]{Gallapreprint}
\bibinfo{author}{\bibfnamefont{T.}~\bibnamefont{Galla}},
  \bibinfo{journal}{preprint arXiv:0711.0169}  (\bibinfo{year}{2007}).

\bibitem[{\citenamefont{Dorogovtsev and
  Mendes}(2003)}]{Dorogovtsev_Mendes_2003}
\bibinfo{editor}{\bibfnamefont{S.~N.} \bibnamefont{Dorogovtsev}}
  \bibnamefont{and} \bibinfo{editor}{\bibfnamefont{J.~F.~F.}
  \bibnamefont{Mendes}}, eds., \emph{\bibinfo{title}{{E}volution of {N}etworks:
  {F}rom {B}iological {N}ets to the {I}nternet and {WWW}}}
  (\bibinfo{publisher}{Oxford Univ. Press}, \bibinfo{year}{2003}).

\bibitem[{\citenamefont{Dunne et~al.}(2002)\citenamefont{Dunne, Williams, and
  Martinez}}]{Dunne_etal_2002a}
\bibinfo{author}{\bibfnamefont{J.~A.} \bibnamefont{Dunne}},
  \bibinfo{author}{\bibfnamefont{R.~J.} \bibnamefont{Williams}},
  \bibnamefont{and} \bibinfo{author}{\bibfnamefont{N.~D.}
  \bibnamefont{Martinez}}, \bibinfo{journal}{PNAS}
  \textbf{\bibinfo{volume}{99}}, \bibinfo{pages}{12917} (\bibinfo{year}{2002}).

\bibitem[{\citenamefont{Hartmann and Weigt}(2005)}]{Hartmann_Weigt_2005}
\bibinfo{author}{\bibfnamefont{A.~K.} \bibnamefont{Hartmann}} \bibnamefont{and}
  \bibinfo{author}{\bibfnamefont{M.}~\bibnamefont{Weigt}},
  \emph{\bibinfo{title}{Phase transitions in combinatorial optimisation
  problems}} (\bibinfo{publisher}{Wiley-VCH}, \bibinfo{year}{2005}).

\bibitem[{\citenamefont{Coolen et~al.}(2005)\citenamefont{Coolen, Skantzos,
  Castillo, Vicente, Hatchett, Wemmenhove, and
  Nikoletopoulos}}]{Coolen_et_al_2005}
\bibinfo{author}{\bibfnamefont{A.~C.~C.} \bibnamefont{Coolen}},
  \bibinfo{author}{\bibfnamefont{N.~S.} \bibnamefont{Skantzos}},
  \bibinfo{author}{\bibfnamefont{I.~P.} \bibnamefont{Castillo}},
  \bibinfo{author}{\bibfnamefont{C.~J.} \bibnamefont{Perez Vicente}},
  \bibinfo{author}{\bibfnamefont{J.~P.~L.} \bibnamefont{Hatchett}},
  \bibinfo{author}{\bibfnamefont{B.}~\bibnamefont{Wemmenhove}},
  \bibnamefont{and}
  \bibinfo{author}{\bibfnamefont{T.}~\bibnamefont{Nikoletopoulos}},
  \bibinfo{journal}{J. Phys. A: Math. Gen.} \textbf{\bibinfo{volume}{38}},
  \bibinfo{pages}{8289} (\bibinfo{year}{2005}).

\bibitem[{\citenamefont{McKane and Newman}(2004)\citenamefont{McKane, and
  Newman}}]{McKane_Newman_2004}
\bibinfo{author}{\bibfnamefont{A.~J.}~\bibnamefont{McKane}},
  \bibnamefont{and}
  \bibinfo{author}{\bibfnamefont{T.~J.} \bibnamefont{Newman}},
  \bibinfo{journal}{Phys. Rev. E} \textbf{\bibinfo{volume}{70}},
  \bibinfo{pages}{041902} (\bibinfo{year}{2004}).

\bibitem[{\citenamefont{McKane and Newman}(2005)\citenamefont{McKane, and
  Newman}}]{McKane_Newman_2005}
\bibinfo{author}{\bibfnamefont{A.~J.}~\bibnamefont{McKane}},
  \bibnamefont{and}
  \bibinfo{author}{\bibfnamefont{T.~J.} \bibnamefont{Newman}},
  \bibinfo{journal}{Phys. Rev. Lett.} \textbf{\bibinfo{volume}{94}},
  \bibinfo{pages}{218102} (\bibinfo{year}{2005}).

\bibitem[{\citenamefont{Coolen}(2000)}]{Coolen_2000}
\bibinfo{author}{\bibfnamefont{A.~C.~C.}~\bibnamefont{Coolen}},
  \emph{\bibinfo{title}{{H}andbook of {B}iological {P}hysics vol. 4, ed F Moss and S Gielen}} (\bibinfo{publisher}{Elsevier},
  \bibinfo{address}{Amsterdam}, \bibinfo{year}{2000}).

\bibitem[{\citenamefont{Eissfeller and Opper}(1992)}]{Eissfeller_Opper_1992}
\bibinfo{author}{\bibfnamefont{H.} \bibnamefont{Eissfeller}} \bibnamefont{and}
  \bibinfo{author}{\bibfnamefont{M.} \bibnamefont{Opper}},
  \bibinfo{journal}{Phys. Rev. Lett.} \textbf{\bibinfo{volume}{68}},
  \bibinfo{pages}{2094} (\bibinfo{year}{1992}).


\end{thebibliography}

\end{document}